\documentclass[aps,prd,%multicol,twocolumn, 
preprintnumbers,groupedaddress]{revtex4}
\usepackage{epsbox}
\usepackage{amsmath}
\usepackage{amsmath,amssymb}
\usepackage[dvips]{graphicx}

\begin{document}
\preprint{\vtop{
{\hbox{YITP-11-68}\vskip-0pt
                 \hbox{KANAZAWA-11-12} \vskip-0pt
%                 \hbox{hep-ph/07????} 
}
}
}

%\date{\today}

\title{
$\bf{X(3872)}$ and Its Iso-Triplet Partners
}

\author{
Kunihiko Terasaki   %authors' name%
}
\affiliation{
Yukawa Institute for Theoretical Physics, Kyoto University,
Kyoto 606-8502, Japan \\
Institute for Theoretical Physics, Kanazawa University, 
Kanazawa 920-1192, Japan
}

\begin{abstract}
{
Decays of $X(3872)$ and its partners as hidden-charm axial-vector tetra-quark mesons 
are studied. As the result, it is seen that the iso-triplet partners of $X(3872)$ can be 
broad, and therefore, higher statistics will be needed to find them. 
}
\end{abstract}

\maketitle

Tetra-quark mesons are classified into the following four 
groups        
%%%%%%%%%%%%%%%%%%%%%%%%%%%%%%%%%%%%%%%%%%%%%%%%%%%%%%%%%%%%%%%%%%%%%%%%
\begin{eqnarray} 
&&\hspace{-8mm} \{qq\bar q\bar q\} =  
[qq][\bar q\bar q] \oplus (qq)(\bar q\bar q)  
\oplus \{[qq](\bar q\bar q)\oplus (qq)[\bar q\bar q]\},\quad(q = u,d,s, c) 
                                                   \label{eq:4-quark} 
\end{eqnarray} 
%%%%%%%%%%%%%%%%%%%%%%%%%%%%%%%%%%%%%%%%%%%%%%%%%%%%%%%%%%%%%%%%%%%%%%%%
in a quark model~\cite{D_{s0}-KT} as in the MIT bag 
model~\cite{Jaffe}, where parentheses and square brackets denote symmetry 
and anti-symmetry, respectively, of flavor wave functions under exchange of flavors 
between them. 
Each term in the right-hand-side of Eq.~(\ref{eq:4-quark}) is again classified into two 
groups~\cite{Jaffe} with  
%%%%%%%%%%%%%%%%%%%%%%%%%%%%%%%%%%%%%%%%%%%%%%%%%%%%%%%%%%%%%%%%%%%%%%%%
${\bf \bar 3_c}\times{\bf 3_c}$ and ${\bf 6_c}\times {\bf \bar 6_c}$  
%%%%%%%%%%%%%%%%%%%%%%%%%%%%%%%%%%%%%%%%%%%%%%%%%%%%%%%%%%%%%%%%%%%%%%%%
of the color $SU_c(3)$. 
However,  the former is expected to be lower in the case of heavy mesons, because 
forces~\cite{color} between two quarks are attractive when they are of 
${\bf \bar{3}_c}$, while repulsive when ${\bf 6_c}$, and because a possible mixing 
between these two states is expected to be small at the scale of heavy meson 
mass~\cite{HT-isospin}. 
Regarding with the spin ($J$),  the $[qq]$ and $(qq)$ have $J=0$ and $1$, respectively, 
and hence the spin and parity $J^P$ of $[qq][\bar q\bar q]$ and 
$\{[qq](\bar q\bar q)\oplus (qq)[\bar q\bar q]\}$ are $J^P=0^+$ and $1^+$, respectively, 
in the flavor symmetry limit. 
In the real world, however, the flavor symmetry is broken and hence the above 
tetra-quark states can have $J=0,\,1$ and $2$, in general.  
Nevertheless, no indication of tetra-quark meson with $J^P=2^+$ has been 
observed, so that the above assignment seems to be favored in the real world. 
Thus, we treat the $[qq][\bar q\bar q]$ mesons as scalar ones~\cite{D_{s0}-KT} 
and  the $\{[qq](\bar q\bar q)\oplus (qq)[\bar q\bar q]\}$ as axial-vector 
ones~\cite{X-3872-KT}, in contrast with Ref.~\cite{Maiani-X} in which 
$X(3872)$ has been assigned to $[cn][\bar {c}\bar {n}]$ with a large violation of 
isospin symmetry. 
The charm strange scalar $D_{s0}^+(2317)$ observed in $e^+e^-$ 
annihilation~\cite{D_{s0}-e^+e^-} as well as in $B$ decays~\cite{Belle-D_{s0}} and 
the $\eta\pi^0$ peak at $3.2$ GeV in two photon 
collision~\cite{hidden-charm-scalar-Belle}, respectively, are good 
candidates~\cite{D_{s0}-KT,HT-isospin,hidden-charm-scalar-KT} of  
$[cn][\bar s\bar n]_{I=1}\sim \hat{F}_I$ and 
$[cn][\bar c\bar n]_{I=1}\sim \hat{\delta}^{c}$, ($n=u,\,d$).  
In addition, an indication of another charm strange scalar which has been observed 
in the $D_s^{*+}\gamma$ channel of $B$ decays~\cite{Belle-D_{s0}} is a suitable   
candidate of the iso-singlet partner~\cite{D_{s0}-KT,HT-isospin} 
$[cn][\bar s\bar n]_{I=0}\sim \hat{F}_0^+$ of $D_{s0}^+(2317) = \hat{F}_I^+$. 
However, the hidden charm~\cite{Belle-X(3872)} $X(3872)$ has a mass much higher 
than that of the above candidate of  the hidden-charm scalar $\hat{\delta}^{c}$, and 
its spin-parity is favored to be $J^P=1^+$ or $2^-$ by 
experiments~\cite{Belle-X-JP,Babar-X-JP}. 
Therefore, we assign~\cite{X-3872-KT}  $X(3872)$ to 
$\{[cn](\bar c\bar n) + (cn)[\bar c\bar n]\}$ (but not to $[cn][\bar c\bar n]$) with 
$J^P=1^+$. 
However, we ignore~\cite{NFQCD} the $(qq)({\bar q}{\bar q})$ mesons in this short 
note. 

Now we review very briefly $X(3872)$ for later discussions. 
A recent analysis~\cite{most-recent} in 
$X(3872) \rightarrow \pi^+ \pi^- J/\psi$ provides its mass and width as  
%%%%%%%%%%%%%%%%%%%%%%%%%%%%%%%%%%%%%%%%%%%%%%%%%%%%%%%%%%%%%%%%%%%%%%%%
\begin{equation}m_{X} = 3871.56 \pm 0.22\,\, {\rm MeV} \,\,{\rm and}\,\,
\Gamma_{X} < 1.2 \,\,{\rm MeV} \,\,(90 \% {\rm CL}). 
                                                                          \label{eq:fitted-width-X-3872}
\end{equation}
%%%%%%%%%%%%%%%%%%%%%%%%%%%%%%%%%%%%%%%%%%%%%%%%%%%%%%%%%%%%%%%%%%%%%%%%
Because it decays into $\gamma J/\psi$ state, its charge conjugation parity 
($\mathcal{C}$) is even~\cite{PDG10}, and hence, we here take~\cite{Belle-X-JP} 
$J^{P\mathcal{C}} = 1^{++}$. 
Although the $X(3872)\rightarrow\pi^+\pi^-J/\psi$ decay proceeds through the 
$\rho^0J/\psi$ intermediate state~\cite{Belle-X-gamma-psi,CDF-X-rho}, its isospin 
quantum number is favored to be $I=0$, because no indication of charged partner of 
$X(3872)$ has been observed~\cite{Babar-X_I-search}. 
In addition, the above isospin assignment is consistent with the observations of the 
$X(3872)\rightarrow\omega J/\psi\rightarrow\pi^+\pi^-\pi^0J/\psi$ 
decay~\cite{Belle-X-gamma-psi,Babar-X-JP}.  
Thus the $X(3872)\rightarrow\pi^+\pi^-J/\psi$ decay is isospin non-conserving.  
Here, it should be noted that the rate for the $X(3872)\rightarrow\pi^+\pi^-J/\psi$ 
decay is nearly equal~\cite{Choi} to that for the 
$X(3872)\rightarrow \pi^+\pi^-\pi^0J/\psi$,  
%%%%%%%%%%%%%%%%%%%%%%%%%%%%%%%%%%%%%%%%%%%%%%%%%%%%%%%%%%%%%%%%%%%%%%%%
\begin{equation}
%R\equiv 
\frac{Br(X(3872)\rightarrow \pi^+\pi^-\pi^0J/\psi)}
{Br(X(3872)\rightarrow \pi^+\pi^-J/\psi)}
=0.8\pm 0.3.                                                                 \label{eq:3pi/2pi}
\end{equation}
%%%%%%%%%%%%%%%%%%%%%%%%%%%%%%%%%%%%%%%%%%%%%%%%%%%%%%%%%%%%%%%%%%%%%%%%
Comparing the above ratio with the measured ratio 
$\displaystyle{{\Gamma(\omega\rightarrow 3\pi)}/
{\Gamma(\omega\rightarrow 2\pi)}\sim 60}$, 
one might feel that  Eq.~(\ref{eq:3pi/2pi}) is strange. 
However, these two ratios are not necessarily parallel to each other, as seen below. 
The $X(3872)\rightarrow \omega J/\psi\rightarrow \rho^0J/\psi\rightarrow 
\pi^+\pi^-J/\psi$ 
decay in the denominator of Eq.~(\ref{eq:3pi/2pi}) is extraordinarily enhanced, 
because of double pole contribution of $\omega$ and $\rho^0$ with 
$m_\omega^2 - m_\rho^2 \ll m_\omega^2$, where the broad width~\cite{PDG10} 
$\Gamma_\rho\simeq 150$ MeV of  $\rho$ has to be taken into account, and 
because the kinematical condition of the former in which $\omega$ decays into 
$\pi^+\pi^-\pi^0$ in the energy region lower than 
$m_{X(3872)} - m_{J/\psi} \simeq m_\omega - \Gamma_\omega/2$ is different from 
that of the latter (on the mass-shell of $\omega$), so that the rate for 
the $X(3872)\rightarrow \pi^+\pi^-\pi^0J/\psi$ decay might be sensitive to a 
mechanism of $\omega\rightarrow 3\pi$ and hence that of 
$X(3872)\rightarrow \pi^+\pi^-\pi^0J/\psi$. 
Nevertheless, the mechanism of $\omega\rightarrow 3\pi$ is still uncertain. 
To see this, we consider the $\omega\rightarrow 3\pi$ decay and the radiative 
$\omega\rightarrow \gamma\pi^0$ in addition to 
$\rho^{\pm,0}\rightarrow \gamma\pi^{\pm,0}$ under the vector meson 
dominance~\cite{VMD-Terasaki} (VMD). 
By taking the measured rate~\cite{PDG10}  
$\Gamma(\omega\rightarrow\gamma\pi^0)_{\rm exp} = 701\pm 25$ keV as the input 
data, our calculated rate, 
$\Gamma(\rho\rightarrow\gamma\pi)_{\rm th} \simeq 72 - 73$ keV reproduces 
considerably well the measured rates~\cite{PDG10}, 
$\Gamma(\rho^\pm\rightarrow\gamma\pi^\pm)_{\rm exp} \simeq 67\pm 8$ keV 
and 
$\Gamma(\rho^0\rightarrow\gamma\pi^0)_{\rm exp} \simeq 90\pm 12$ keV,   
although the measured rates still have large ambiguities. 
From the above, it is seen that the VMD works in these decays, at least in 
the $\omega\rightarrow\gamma\pi^0$ and $\rho^\pm\rightarrow\gamma\pi^\pm$ 
decays. 
Next, we determine the $\omega\rho\pi$ coupling strength from the above 
$\Gamma(\omega\rightarrow\gamma\pi^0)_{\rm exp}$ and apply it to the 
$\omega\rightarrow\rho\pi\rightarrow 3\pi$ decay. 
However the resulting rate 
$\Gamma(\omega\rightarrow\rho\pi\rightarrow 3\pi)_{\rm th}\simeq 5$ MeV 
fails to reproduce the measured one~\cite{PDG10}, 
$\Gamma(\omega\rightarrow 3\pi)_{\rm exp} = 7.57\pm 0.09$ MeV. 
It suggests that some extra contribution(s) are needed, because the contribution of 
$\rho$ meson pole is sizable but insufficient, i.e., the mechanism of the 
$\omega\rightarrow 3\pi$ decay and hence that of the 
$X(3872)\rightarrow\omega J/\psi\rightarrow \pi^+\pi^-\pi^0 J/\psi$ 
are still uncertain and not simple. 
For this reason, we have considered the $X(3872)\rightarrow \gamma J/\psi$ decay 
in place of the $X(3872)\rightarrow \pi^+\pi^-\pi^0 J/\psi$ in 
Ref.~\cite{omega-rho-KT}. 
Although the measured ratio~\cite{Belle-X-gamma-psi,Babar-gamma-psi'} of the rates 
$\Gamma(X(3872)\rightarrow \gamma J/\psi)
/\Gamma(X(3872)\rightarrow \pi^+\pi^- J/\psi)$ 
is less than unity against the well-known hierarchy of hadron 
interactions~\cite{HT-isospin}, 
%%%%%%%%%%%%%%%%%%%%%%%%%%%%%%%%%%%%%%%%%%%%%%%%%%%%%%%%%%%%%%%%%%%%%%%%
$|${\it isospin conserving int.} ($\sim O(1)$)$|$ $\gg$ 
$|${\it electromagnetic int.} ($\sim O(\sqrt{\alpha})$)$|$
$\gg$ $|${\it isospin non-conserving int.}
($\sim O(\alpha)$~\cite{Dalitz})$|$,                                                             
%%%%%%%%%%%%%%%%%%%%%%%%%%%%%%%%%%%%%%%%%%%%%%%%%%%%%%%%%%%%%%%%%%%%%%%%
it has been 
reproduced~\cite{omega-rho-KT,NFQCD} in our scheme that $X(3872)$ 
is the tetra-quark meson given before and the $\omega\rho^0$ mixing is the origin 
of the isospin non-conservation. 
It suggests again that the $X(3872)\rightarrow \pi^+\pi^- J/\psi$ is extraordinarily 
enhanced. 
In contrast, if $X(3872)$ were a charmonium, the ratio of decay rates under 
consideration could not be reproduced~\cite{omega-rho-KT,NFQCD}, 
because such an enhancement cannot work in this case. 

Analyses in the $D^0\bar{D}^{*0} + c.c. \,\,(\rightarrow D^0\bar{D}^0\pi^0$ and 
$D^0\bar{ D}^0\gamma)$ channels also have reported observations~\cite{X-3875-exp}  
of $X(3875)$.  
Recent results~\cite{Belle-D^0D^{*0}bar} on its mass and width are 
%%%%%%%%%%%%%%%%%%%%%%%%%%%%%%%%%%%%%%%%%%%%%%%%%%%%%%%%%%%%%%%%%%%%%%%%
\begin{equation}
m_{X(3875)} = 3872.6^{+0.6+0.4}_{-0.4- 0.5}\,\, 
{\rm MeV} \,\,{\rm and}\,\,  
\Gamma_{X(3875)} = 3.9^{+2.8+0.2}_{-1.4-1.1}\,\, 
{\rm MeV}.                                                                \label{eq:most-recent}
\end{equation} 
%%%%%%%%%%%%%%%%%%%%%%%%%%%%%%%%%%%%%%%%%%%%%%%%%%%%%%%%%%%%%%%%%%%%%%%%
If the numerical results in Eqs.~(\ref{eq:fitted-width-X-3872}) and 
(\ref{eq:most-recent}) were literally accepted, $X(3875)$ and $X(3872)$ would be 
different states. 
However,  it is unnatural to assign $X(3875)$ and $X(3872)$ to different states as 
will be discussed later, and therefore,  we here presume that the narrow $X(3875)$ 
and $X(3872)$ are identical. 
In this case, the averaged ratio of rates for the 
$X(3872) = X(3875)\rightarrow D^0\bar{D}^{*0} + c.c.$ decay to the 
$X(3872)\rightarrow \pi^+\pi^-J/\psi$ has been given by~\cite{Choi}  
%%%%%%%%%%%%%%%%%%%%%%%%%%%%%%%%%%%%%%%%%%%%%%%%%%%%%%%%%%%%%%%%%%%%%%%%
\begin{equation}
%R(D^0\bar{D}^{*0}) = 
\frac{\Gamma(X(3872)\rightarrow D^0\bar{D}^{*0} + c.c.)}
{\Gamma(X(3872)\rightarrow \pi^+\pi^-J/\psi)} = 9.5\pm 3.1.
                                                                                 \label{eq:R-D^0barD^{*0}}
\end{equation}
%%%%%%%%%%%%%%%%%%%%%%%%%%%%%%%%%%%%%%%%%%%%%%%%%%%%%%%%%%%%%%%%%%%%%%%%

Now, we study possible decay modes of $X(3872)$ and its partners. 
In the  present scheme, hidden-charm iso-singlet axial-vector tetra-quark mesons 
with $\mathcal{C} = \pm$ are given by $X(\pm) = \{X_u(\pm) + X_d(\pm)\}/\sqrt{2}$, 
where $X_u(\pm)$ and $X_d(\pm)$ are provided by 
%%%%%%%%%%%%%%%%%%%%%%%%%%%%%%%%%%%%%%%%%%%%%%%%%%%%%%%%%%%%%%%%%%%%%%%%
\begin{eqnarray}
&&
X_u(\pm) 
= \displaystyle{\frac{1}{2\sqrt{2}}\Bigl\{[cu]_{\bar 3_c}^{1_s}(\bar c\bar u)_{3_c}^{3_s} 
\pm (cu)_{\bar 3_c}^{3_s}[\bar c\bar u]_{3_c}^{1_s} \Bigr\}_{1_c}^{3_s}},
                                                                                       \label{eq:X_u-def} \\
&&
X_d(\pm) 
= \displaystyle{\frac{1}{2\sqrt{2}}\Bigl\{[cd]_{\bar 3_c}^{1_s}(\bar c\bar d)_{3_c}^{3_s} 
\pm (cd)_{\bar 3_c}^{3_s}[\bar c\bar d]_{3_c}^{1_s}\Bigr\}_{1_c}^{3_s}}.
                                                                                          \label{eq:X_d-def}
\end{eqnarray}  
%%%%%%%%%%%%%%%%%%%%%%%%%%%%%%%%%%%%%%%%%%%%%%%%%%%%%%%%%%%%%%%%%%%%%%%%
Here the subscripts $1_c$,  $\bar 3_c$,  $3_c$ denote the color multiplets, and the 
superscripts $1_s$ and $3_s$ the spin multiplets. %, i.e., $J =0$ and $1$, respectively.  
The above $X_u(\pm)$ can be decomposed as 
%%%%%%%%%%%%%%%%%%%%%%%%%%%%%%%%%%%%%%%%%%%%%%%%%%%%%%%%%%%%%%%%%%%%%%%%
\begin{eqnarray}
&&X_u(+) 
= \displaystyle{\frac{1}{2}
\sqrt{\frac{1}{6}}\bigl\{\sqrt{2}(c\bar c)_{1_c}^{3_s}(u\bar u)_{1_c}^{3_s}
+ (c\bar u)_{1_c}^{1_s}(u\bar c)_{1_c}^{3_s}   
+ (c\bar u)_{1_c}^{3_s}(u\bar c)_{1_c}^{1_s}   
\bigr\}_{1_c}^{3_s} + \cdots} 
\nonumber\\
&& \hspace{15mm}
 - \displaystyle{\frac{1}{2}
\sqrt{\frac{1}{6}}\bigl\{(u\bar c)_{1_c}^{1_s}(c\bar u)_{1_c}^{3_s}    
+ (u\bar c)_{1_c}^{3_s}(c\bar u)_{1_c}^{1_s}        
+ \sqrt{2} (u\bar u)_{1_c}^{3_s}(c\bar c)_{1_c}^{3_s}      
\bigr\}_{1_c}^{3_s} + \cdots},                                                     \label{eq:X_u(+)}      
\\
&&
X_u(-) 
= \displaystyle{\frac{1}{2}
\sqrt{\frac{1}{6}}\bigl\{(c\bar c)_{1_c}^{1_s}(u\bar u)_{1_c}^{3_s} 
+ (c\bar c)_{1_c}^{3_s}(u\bar u)_{1_c}^{1_s}                           %\rangle_{1_c}^{3_s}
+ \sqrt{2}(c\bar u)_{1_c}^{3_s}(u\bar c)_{1_c}^{3_s}}\bigr\}_{1_c}^{3_s} + \cdots
\nonumber
\\
&&\hspace{15mm}
\displaystyle{-\frac{1}{2}
\sqrt{\frac{1}{6}}\bigl\{\sqrt{2}(u\bar c)_{1_c}^{3_s}(c\bar u)_{1_c}^{3_s}    %\rangle 
+ (u\bar u)_{1_c}^{1_s}(c\bar c)_{1_c}^{3_s}      %\rangle_{1_c}^{3_s}
+ (u\bar u)_{1_c}^{3_s}(c\bar c)_{1_c}^{1_s}     %\rangle_{1_c}^{1_s}
\bigr\}_{1_c}^{3_s} + \cdots.                             
}
                                                                                       \label{eq:X_d(+)}      
\end{eqnarray}
%%%%%%%%%%%%%%%%%%%%%%%%%%%%%%%%%%%%%%%%%%%%%%%%%%%%%%%%%%%%%%%%%%%%%%%%
where $\cdots$ denotes a color-singlet sum of products of color-octet $\{q\bar q\}$ 
pairs.  
Decompositions of $X_d(\pm)$ are obtained by replacing $u$ by $d$ in the above 
equations. 
Replacement of the color singlet $\{q\bar{q}\}$ pairs, 
$\{u\bar u - d\bar d\}_{1_c}^{1_s}/\sqrt{2}$, 
$\{u\bar u + d\bar d\}_{1_c}^{1_s}/\sqrt{2}$, 
$\{c\bar c\}_{1_c}^{1_s}$, etc. 
by the ordinary $\pi^0$, $\eta_0$, $\eta_c$, etc., respectively, leads to 
%%%%%%%%%%%%%%%%%%%%%%%%%%%%%%%%%%%%%%%%%%%%%%%%%%%%%%%%%%%%%%%%%%%%%%%%
\begin{eqnarray}
&& 
X(+) = \displaystyle{\frac{1}{4}\sqrt{\frac{1}{3}}\bigl\{
2(J/\psi\omega - \omega J/\psi) + (D^{0}\bar D^{*0} - \bar D^{*0} D^{0}) +  
(D^{*0}{\bar  D}^{0} - {\bar D}^{0}D^{*0})}
\nonumber\\
&&\hspace{30mm}
+ (D^{+}\bar D^{*-} - D^{*-}D^+) + (D^{*+}D^- - D^-D^{*+})\bigr\}  + \cdots,               
                                                                                      \label{eq:decomp_X(+)}      
\\
&&
X(-) = \displaystyle{\frac{1}{4}\sqrt{\frac{2}{3}}\bigl\{
(\eta_c\omega - \omega\eta_c) + (J/\psi\eta_0 - \eta_0J/\psi)} 
\nonumber\\
&&\hspace{30mm}
+ (D^{*0}\bar D^{*0} - \bar D^{*0}D^{*0}) + (D^{*+}D^{*-} - D^{*-}D^{*+})
\bigr\} + \cdots.
                                                                                  \label{eq:decomp_X(-)} 
\end{eqnarray}
%%%%%%%%%%%%%%%%%%%%%%%%%%%%%%%%%%%%%%%%%%%%%%%%%%%%%%%%%%%%%%%%%%%%%%%%
Their iso-triplet neutral partners $X^0_I(\pm)$ also can be decomposed as 
%%%%%%%%%%%%%%%%%%%%%%%%%%%%%%%%%%%%%%%%%%%%%%%%%%%%%%%%%%%%%%%%%%%%%%%%
\begin{eqnarray}
&& 
X^0_I(+) = \displaystyle{\frac{1}{4}\sqrt{\frac{1}{3}}\bigl\{
2(J/\psi\rho^0 - \rho^0 J/\psi) + (D^{0}\bar D^{*0} - \bar D^{*0} D^{0}) + 
({\bar D}^{0}D^{*0} - D^{*0}{\bar  D}^{0})}
\nonumber\\
&&\hspace{25mm}
- (D^{+}\bar D^{*-} - D^{*-}D^+) - (D^{*+}D^- - D^-D^{*+})\bigr\}  + \cdots,               
                                                                                \label{eq:decomp_X_I(+)}      
\\
&&
X^0_I(-) = \displaystyle{\frac{1}{4}\sqrt{\frac{2}{3}}\bigl\{
(\eta_c\rho^0 - \rho^0\eta_c) + (J/\psi\pi^0 - \pi^0J/\psi)} 
\nonumber\\
&&\hspace{25mm}
+ (D^{*0}{\bar D}^{*0} - {\bar D}^{*0}D^{*0}) - (D^{*+}D^{*-} - D^{*-}D^{*+})
\bigr\} + \cdots.                                                               \label{eq:decomp_X_I(-)} 
\end{eqnarray}
%%%%%%%%%%%%%%%%%%%%%%%%%%%%%%%%%%%%%%%%%%%%%%%%%%%%%%%%%%%%%%%%%%%%%%%% 
From Eqs.~(\ref{eq:decomp_X(+)}) -- (\ref{eq:decomp_X_I(-)}), we can see 
(i) $X(+)$, to which $X(3872)$ is assigned, couples to $\omega J/\psi$ and 
$D^0\bar{D}^{*0} + c.c.$ 
Therefore, it can decay into $\pi^+\pi^-\pi^0J/\psi$ (and $\pi^+\pi^-J/\psi$) 
through $\omega$ pole  (with the $\omega\rho^0$ mixing) and into 
$D^0\bar{D}^0\pi^0\,({\rm or}\,\,\gamma)$ through $D^0\bar{D}^{*0} + c.c.$, as have 
been observed. 
It is also seen that rates for these decays are small because of small overlap of 
flavor, color and spin wave functions as the 
$\hat{F}_I^+ = D_{s0}^+(2317)\rightarrow D_s^+\pi^0$ decay\cite{HT-isospin}, 
isospin non-conservation in the $\pi^+\pi^-J/\psi$ decay and very small phase space 
volumes in the $D^0\bar{D}^0\pi^0$ decay through the $D^0\bar{D}^{*0} + c.c.$ and 
the $\pi^+\pi^-\pi^0J/\psi$ decay through the $\omega J/\psi$. 
(ii) Its opposite $\mathcal{C}$-parity partner $X(-)$ couples to $J/\psi\eta_0$, 
$\eta_c\omega$ and $D^{*}\bar{D}^{*}$. 
Nevertheless, the threshold of $D^{*}{\bar{D}}^{*}$ decay is beyond $m_{X(3872)}$, 
and hence probably  beyond $m_{X(-)}$. 
Therefore, $X(-)$ might be observed in the $\eta J/\psi$ channel. 
(iii) The iso-triplet partners $X_I(+)$'s of $X(3872)$ couple to $\rho J/\psi$  and 
$D\bar{D}^{*} $ (and $\bar{D}{D}^{*}$). 
Therefore, one might identify $X^0_I(+)$ with $X(3875)$ observed in the 
$D^0\bar{D}^{*0} + c.c.$ channel. 
However, it should be noted that the rate for the  isospin conserving 
$X^0_I(+)\rightarrow\rho^0J/\psi\rightarrow \pi^+\pi^-J/\psi$ decay 
will be much larger than that for the isospin non-conserving 
$X(3872)\rightarrow\omega J/\psi\rightarrow\rho^0J/\psi
\rightarrow \pi^+\pi^-J/\psi$, 
as will be explicitly seen later. 
Therefore, $X^0_I(+)$ can be broad, and hence it seems to be unnatural to assign the 
narrow $X(3875)$ to the hypothetically broad $X^0_I(+)$, as noted before.  
(iv) The iso-triplet partners $X_I(-)$'s with negative $\mathcal{C}$-parity couple to 
$\pi J/\psi$ and $\eta_c\rho$. 
If the spatial wave function of $X_I(-)$ is not very much different from that of 
$X_I(+)$, the rate for the $X_I(-)\rightarrow \pi J/\psi$ would be much larger than 
that for the $X^0_I(+)\rightarrow\rho^0J/\psi\rightarrow \pi^+\pi^-J/\psi$, 
because of much larger phase space volume. 

We here study the rate for the 
$X^0_I(+)\rightarrow\rho^0J/\psi\rightarrow \pi^+\pi^-J/\psi$ decay to see why 
$X^0_I(+)$ has not been observed. 
In Eq.~(22) of Ref.~\cite{omega-rho-KT}, we have calculated the rate for the 
%%%%%%%%%%%%%%%%%%%%%%%%%%%%%%%%%%%%%%%%%%%%%%%%%%%%%%%%%%%%%%%%%%%%%%%%
$X(3872)\rightarrow \omega J/\psi\rightarrow\rho^0J/\psi
\rightarrow\pi^+\pi^-J/\psi$ 
%%%%%%%%%%%%%%%%%%%%%%%%%%%%%%%%%%%%%%%%%%%%%%%%%%%%%%%%%%%%%%%%%%%%%%%%
decay with the $\omega\rho^0$ mixing. 
The rate for the above decay of $X^0_I(+)$ can be obtained by replacing $X(3872)$ by 
$X^0_I(+)$ and eliminating the contribution of $\omega$ pole with the 
$\omega\rho^0$ mixing in the equation. 
Taking $m_{X_I(+)} \simeq m_{X(3872)}$ (because both of them consist of the same 
quarks and their flavor wave functions are of the same type, as in the case of 
$\hat{F}_I^+ = D_{s0}^+(2317)$ and $\hat{F}_0^+$ which have been observed as signal 
and indication, respectively, at the same mass in $B$ decays~\cite{Belle-D_{s0}} as 
discussed before), we get the ratio of rates 
%%%%%%%%%%%%%%%%%%%%%%%%%%%%%%%%%%%%%%%%%%%%%%%%%%%%%%%%%%%%%%%%%%%%%%%%
\begin{equation}
\frac{\Gamma(X^0_I(+)\rightarrow\rho^0J/\psi\rightarrow \pi^+\pi^-J/\psi)}
{\Gamma(X(3872)                   %\rightarrow\omega J/\psi\rightarrow\rho^0J/\psi
\rightarrow \pi^+\pi^-J/\psi)}
\sim 200,                                                          \label{eq:ratio-of-rates}
\end{equation}
%%%%%%%%%%%%%%%%%%%%%%%%%%%%%%%%%%%%%%%%%%%%%%%%%%%%%%%%%%%%%%%%%%%%%%%%
where it has been assumed that the size of the $X^0_I(+)\rho^0J/\psi$ coupling is 
approximately equal to that of the $X(+)\omega J/\psi$, because the spatial wave 
functions of  $X_I(+)$ and  $X(+)$ are expected to be not very much different from 
each other, as in the case~\cite{HT-isospin} of $\hat{F}_I^+$ and $\hat{F}_0^+$. 

To estimate the denominator of Eq.~(\ref{eq:ratio-of-rates}), we assume that the full 
width of $X(3872)$ is approximately saturated as 
%%%%%%%%%%%%%%%%%%%%%%%%%%%%%%%%%%%%%%%%%%%%%%%%%%%%%%%%%%%%%%%%%%%%%%%%
\begin{eqnarray}
&&\Gamma_{X(3872)} \simeq \Gamma(X(3872) \rightarrow \pi^+\pi^-J/\psi
+ \Gamma(X(3872) \rightarrow \pi^+\pi^-\pi^0J/\psi) \nonumber\\
&&\hspace{19mm}
+ \Gamma(X(3872)\rightarrow D^0\bar{D}^{*0} + c.c.). 
\end{eqnarray}
%%%%%%%%%%%%%%%%%%%%%%%%%%%%%%%%%%%%%%%%%%%%%%%%%%%%%%%%%%%%%%%%%%%%%%%%
By using Eqs.~(\ref{eq:3pi/2pi}) and (\ref{eq:R-D^0barD^{*0}}),  
$\Gamma(X(3872)\rightarrow \pi^+\pi^-J/\psi)$ can be given by $\Gamma_{X(3872)}$. 
We have listed two different values, $\Gamma_{X(3872)}$ in 
Eq.~(\ref{eq:fitted-width-X-3872}) and $\Gamma_{X(3875)}$ in 
Eq.~(\ref{eq:most-recent}), where $X(3872)$ and $X(3875)$ are now identified. 
The latter is consistent with the measured width~\cite{Belle-X(3872)} $2.5\pm 0.5$ 
MeV of $X(3872)$.  
However, this is narrower than the experimental energy resolution, and 
therefore, some corrections might be needed. 
Such a corrected width has been given in Eq.~(\ref{eq:fitted-width-X-3872}). 

Taking $\Gamma_{X(3875)}$ in Eq.~(\ref{eq:most-recent}),  as an example, we obtain 
%%%%%%%%%%%%%%%%%%%%%%%%%%%%%%%%%%%%%%%%%%%%%%%%%%%%%%%%%%%%%%%%%%%%%%%%
$0.1 \lesssim \Gamma(X(3872)   %\rightarrow\omega J/\psi\rightarrow\rho^0J/\psi
\rightarrow \pi^+\pi^-J/\psi) \lesssim 0.9\,\,{\rm MeV}$, 
%%%%%%%%%%%%%%%%%%%%%%%%%%%%%%%%%%%%%%%%%%%%%%%%%%%%%%%%%%%%%%%%%%%%%%%%
and therefore, 
%%%%%%%%%%%%%%%%%%%%%%%%%%%%%%%%%%%%%%%%%%%%%%%%%%%%%%%%%%%%%%%%%%%%%%%%
\begin{equation}
\Gamma(X^0_I(+)\rightarrow\rho^0J/\psi\rightarrow \pi^+\pi^-J/\psi)
\sim (20 - 200) \,\,{\rm MeV}. 
\end{equation}
%%%%%%%%%%%%%%%%%%%%%%%%%%%%%%%%%%%%%%%%%%%%%%%%%%%%%%%%%%%%%%%%%%%%%%%%
The above rate is much larger than that for the near threshold 
$X_I(+)\rightarrow D^0\bar{D}^{*0} + c.c.$ decay, because 
$\Gamma(X_I(+)\rightarrow D^0\bar{D}^{*0} + c.c.)
\simeq \Gamma(X(+)\rightarrow D^0\bar{D}^{*0} + c.c.)$ will be obtained, if 
$m_{X_I(+)} \simeq m_{X(3872)}$ and 
$|g_{X_I(+)D^0\bar{D}^{*0}}|\simeq |g_{X(+)D^0\bar{D}^{*0}}|$ as discussed before. 
% as $|g_{X_I(+)\rho^0J/\psi}|\simeq |g_{X(+)\omega J/\psi}|$.  
Therefore, the full width of $X_I(+)$ would be dominated by 
$\Gamma(X_I(+)\rightarrow\rho J/\psi\rightarrow\pi\pi J/\psi)$, 
and hence $X_I(+)$ would be much broader than $X(3872)$. 
In this case, it is expected that the broad enhancement of the $\pi^+\pi^-J/\psi$ 
mass distribution from $X^0_I(+)$ would be behind the background of the narrow 
$X(3872)$ peak, unless the production rates of $X_I(+)$'s are much larger than that 
of $X(3872)$.   
Although the existing search for the charged partner of $X(3872)$ mentioned 
before has reported~\cite{Babar-X_I-search} no indication of a narrow 
$\pi^-\pi^0J/\psi$ resonance around the mass of $X(3872)$, this does not 
necessarily exclude existence of iso-triplet partners, because their production rate 
has not been known yet and, in addition, the present statistical accuracy might be 
insufficient to observe the broad $X^0_I(+)$ in the $\pi\pi J/\psi$ mass distribution. 
On the other hand, if $\Gamma_{X(3872)}$ in Eq.~(\ref{eq:fitted-width-X-3872}) as 
another example is taken, a small $\Gamma_{X_{I}(+)}$ would be possible. 
In this case, $X_I(+)$'s could have been observed in the $\pi\pi J/\psi$ channels.  
However, the negative result on the search for the $(\pi\pi)^- J/\psi$ might imply  
that the true width of $X(3872)$ is near the upper bound of $\Gamma_{X(3872)}$ in 
Eq.~(\ref{eq:fitted-width-X-3872}), and therefore $X_I(+)$ would be considerably broad, 
if its production rate is of the same order of magnitude as that of $X(3872)$. 
Therefore, more precise determination of intrinsic widths of  $X(3872)$ and its 
partners in addition to their production rates will provide important informations to 
search for these partners. 

In summary we have studied $X(3872)$ and its partners, assigning these axial-vector 
mesons to $\{[cn](\bar c\bar n)\pm (cn)[\bar c\bar n]\}_{I=1,0}$. 
As the results, we have discussed their possible decay modes, and pointed out 
that the iso-triplet partners of $X(3872)$ can be considerably broad     
and therefore higher statistics will be needed to find them. 
%%%%%%%%%%%%%%%%%%%%%%%%%%%%%%%%%%%%%%%%%%%%%%%%%%%%%%%%%%%%%%%%%%%%%%%
\section*{Acknowledgments}    
The author would like to thank Professor H.~J.~Lipkin  for discussions by which 
this work was motivated. 
He also would like to appreciate Professor K.~Miyabayashi for informing the current 
status of $X(3872)$ experiments. 
%%%%%%%%%%%%%%%%%%%%%%%%%%%%%%%%%%%%%%%%%%%%%%%%%%%%%%%%%%%%%%%%%%%%%%%

%%%%%%%%%%%%%%%%%%%%%%%%%%%%%%%%%%%%%%%%%%%%%%%%%%%%%%%%%%%%%%%%%%%%%%

%\end{references}
%%%%%%%%%%%%%%%%%%%%%%%

%%%%%%%%%%%%%%%%%%%%%%%
\end{document}